# AViTExt: Automatic Video Text Extraction
*A new Approach for video content indexing Application*

Baseem Bouaziz
MIRACL: Multimedia Information systems and Advanced Computing Laboratory
Sfax, Tunisia
Bassem.bouaziz@fsegs.rnu.tn

Tarek Zlitni
MIRACL: Multimedia Information systems and Advanced Computing Laboratory
Sfax, Tunisia
tarek.zlitni@gmail.com

Walid Mahdi
MIRACL: Multimedia Information systems and Advanced Computing Laboratory
Sfax, Tunisia
Walid.mahdi@isimssf.rnu.tn

*Abstract*—**In this paper, we propose a spatial temporal video-text detection technique which proceed in two principal steps: potential text region detection and a filtering process. In the first step we divide dynamically each pair of consecutive video frames into sub block in order to detect change. A significant difference between homologous blocks implies the appearance of an important object which may be a text region. The temporal redundancy is then used to filter these regions and forms an effective text region. The experimentation driven on a variety of video sequences shows the effectiveness of our approach by obtaining a 89,39% as precision rate and 90,19 as recall.**

## I. INTRODUCTION

Content-based video indexing, browsing and retrieval tasks require automatic extraction of descriptive features that are relevant. The typical low level features that are extracted in images or video include measures of color, texture and shape. Although these features can easily be extracted, they do not give a clear idea of the video content. Extracting more descriptive features and higher level entities, for example text [1] or human faces [2], has attracted more and more research interest recently. Text embedded in images and video, especially captions provide brief and important content information, such as the name of player s or speakers, the title, location and date of an event, etc. Thus textual information when embedded within a video are very high level clues for the content-based video indexing and it is generally associated to an important event.

Actually several domains 'general public' are concerned by the utility of video text extraction system such TV on demand, video summarization, video content browsing, video searching, etc…

Text in video is most reliable for users to locate their desired video content. Especially the superimposed text, it is intended to carry important information in video. It is typically generated by video title machines or graphical font generators in studios. If these text occurrences could be detected, segmented, and recognized automatically, they would be a valuable source of high-level semantics for indexing and retrieval. In the field of content-based information retrieval, video text detection and segmentation has drawn much attention of many researchers. Existing text detection methods can be classified into connected component methods [3], texture classification methods [4,5], and edge detection methods [6,7]. The connected component methods and edge detection methods assume that text strokes have a certain contrast against background. And texture classification methods treat text as a kind of texture. The text segmentation methods fall into two groups. One group includes color-based methods, and the other group includes stroke-based methods. The color based methods [8] hold the assumption that the text pixels and non-text pixels are of different colors. And they segment text by threshold. On the other hand, stroke-based methods [9,10,11] employ some filters to enhance stroke-like shapes and then detect strokes according to their density.

Most previous approaches extract texts from individual frames independently; while few methods have exploited the temporal redundancy by using multiple frames [12]. Specifically, frame averaging method [12] has been introduced to reduce the influence of the complex background. With these methods, an enhanced image for segmentation is made from the average/minimum/maximum pixel value that occurs in each location during the frames containing the same text.

Unlike Moving texts like rolling text news, which usually less indicates the video content, most of superimposed texts containing useful description information are static. Therefore our work focuses on static superimposed text. Based on observation, this kind of superimposed text has several characteristics and particularly temporal redundancy which means the same text region lasts than tens frames and fixed position which means the location of the text region remains the same.

Considering these characteristics and as continuity of our last work [13], we propose in this paper a new spatio-temporal approach for non-compressed video text extraction. The integration of the temporal aspect solves both the problem of background complexity and that of false detection. The rest of the paper is organized as follows. Section II details our approach. Simulation results of our experiment are reported in Section III and finally Section VI draws the conclusion.

## II. PROPOSED APPROACH

Our approach operates by the detection of potential text regions followed by a filtering process which validate the effective ones. Indeed, the consecutives video frames are generally similar and if the difference between these frames grows notably, it implies the appearance of new objects which could be text regions. For that, we proceed firstly by identifying the edge map of each frame then we binaries them according the technique described in [14] after we calculate difference of each pair of consecutives video frames. When an important difference is detected a split and merge process is started.

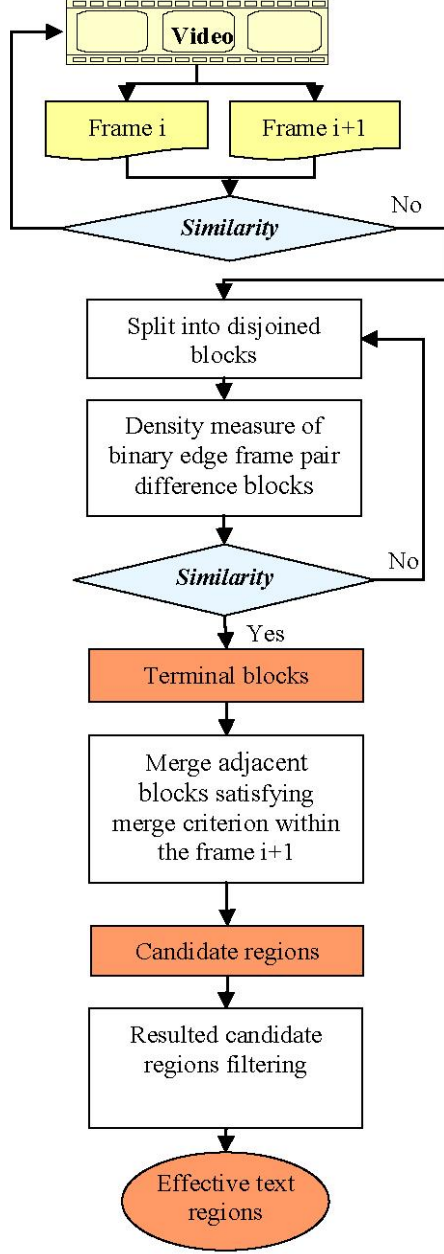

Figure 1. Proposed Approach

The split step consist on subdividing recursively an edge frame in bloc and measuring its edge density and that of its homologous block in the next frame. The subdivision is stopped if the edge density of a bloc is weak or when a bloc is too small (smaller than a readable text). As for the merge step, it consists on merging adjacent blocs having similar edge density. The whole process aims at localizing the new objects which appear within video frame, that's for we need a filtering process which keeps the effective text regions. For that we use two criteria: The first measures locally the contrast of an object with the background and the second evaluate the persistence of the objects in a temporal window.

### A. Candidate regions apperance detection

Most of previous works in the field of video text detection are based on the extraction of spatial features, they applies their algorithms to every N frames, thus the detected regions are issued from each N frames handled independently. Some of these works uses the temporal features only to enhance the quality of detected text.

To reduce the cost of text detection in every frames and to avoid the unnecessary detection we propose in our work a text regions detection which applied only on the frames which may contain text regions.

In order to detect the appearance of new objects we compute the difference between adjacent frame pair according to the HDM measure (Histogram Difference Metric).

$$D_H(i) = \frac{1}{M \times N} \sum_{k=0}^{255} |H_{i+1}(k) - H_i(k)| \quad (1)$$

where $H_i(k)$ denotes the $K$ gray-level value of the $i^{th}$ frame, $M$ and $N$ denotes respectively the width and the height of the frame. The $D_H$ metric is used to determine the appearance of new objects in the frame i+1. If $D_H$ is greater than a threshold $\theta$ then the process of candidate text regions is started otherwise we jump to the next pair of frames.

### B. Candidate Text regions localization

The process of candidate text regions detection aims at detects objects which may be text region. This process is started only if we detect the appearance of new objects in a frame.

Initially, we calculate for each adjacent frame pair the edge map as the square root of the horizontal and vertical Sobel edge map.

$$FE_i = \sqrt{C_X^2 + C_Y^2} \quad (2)$$

Where $FE_i$ $C_X$ and $C_Y$ are respectively $i^{th}$ horizontal and vertical edge frame.

The obtained edge frames pair is binarized according to the optimal thresolding technique [14] which chooses the binarization threshold to minimize the interclass variance of the threshold black and white pixels. This step aims to get a binary image that contains only most contrasted pixels.

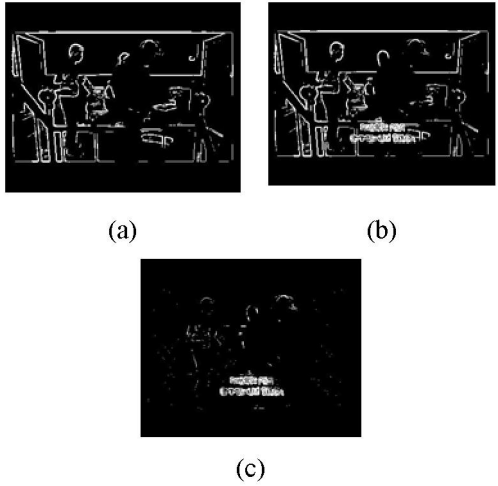

Figure 2. Edge frame pair difference (a) frame i, (b) Frame (i+1), (c) frame difference

After binarization step we calculate the edge frame difference between each pair of frame. Then we proceed by launching a split and merge processes on each binaries edge frame pair according to the quadtree technique [15]. We note that the use of the frame edge difference allows bypassing problems resulting from background complexity and optimize extra processing encountered when used a fixed size and number of blocks.

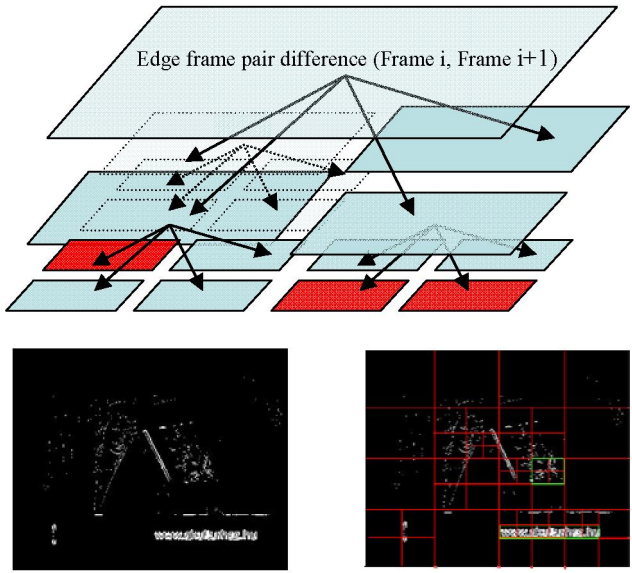

Figure 3. The Split and merge processes by qadtree technique applied on binarised edge frame pair difference

For each bloc we measure the edge density as follow.

$$D = \frac{1}{m \times n} \sum_{x=1}^{m} \sum_{y=1}^{n} I_{edge}(x, y) \qquad (3)$$

Where $I_{edge}(x, y)$ is the intensity value of an edge pixel at the position x, y. $m$ and $n$ denotes respectively the width and the height of the current bloc.

Each block is divided into sub blocks if its edge density D is greater than a predefined threshold T set based on empirical study. The subdivision is stopped only if D is less than T or the bloc is too small. In this case the block become terminal. Two benefits are issued from the use of density measure: the first is to get a dynamic number of split. Thus, block having small edge density value (not contains new objects) are not spitted again. The other benefit is to give information about the region of interest localization since the bloc belonging to desired objects presents generally a high density especially if its text assuming that the colors of pixels belonging to text strokes are identical.

The result of this step is a set of block having a similar pixel density at several level of split. To get the potential text regions the terminal blocks must be grouped within a merge process.

The merge process operates by merging the terminal blocks according to their pixels density and their positions. The adjacent blocks which present similar pixel densities are merged to forms greater blocks and consequently probable text regions. The obtained merged blocks are mapped into the original frame i+1 relative to the current frame pair to get the real regions of interest which will be filtered within a filtering process.

C. *Text region Filtering*

The filtering process is an important task in our approach. The obtained results from the Candidate Text regions localization process constitute a first localization of the regions which seems to be really text. To reduce false detections of the effective text regions, the filtering process aims at validating each already localized region and decides if it is effective or not. For that this filtering process is based on three criteria: the size of region, the contrast variation and the temporal redundancy.

For each candidate text region, we proceed initially by ignoring those having too small size or having width greater than height. This initial filtering step is followed by that of contrast variation. Indeed, when each candidate text region is processed in the way described by optimum thresholding method [14], an analysis of the color spatial variation of all candidate text regions allows us to identify effective text regions. This analysis is based on basic text characteristic which denotes that text characters are generally contrasted with background since artificial text is designed to be read easily. All we need to do is to locate, for each region, pixels belonging to the background and pixels belonging eventually to the text object. Then we determine on the histogram curve of each thresolded region the positions of the two more dominant peaks *P1* and *P2*. If the distance *D(P1, P2)* between the two peaks is greater than a predefined threshold σ, the candidate text region is classified as an effective text region, otherwise it will be ignored. In our experimentation σ = 110.

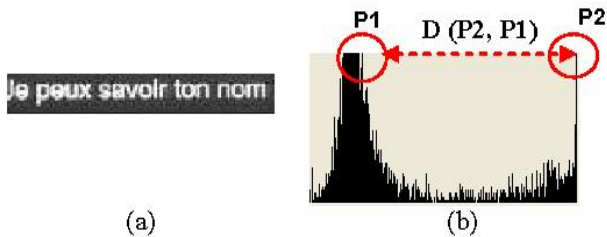

Figure 4. Contrast variation of a text region

However, this filtering criterion fails in the case of regions having a high contrast variation and contains other objects than text. For that, obtained results from this step are also filtered according to their temporal redundancy.

According to the study carried out in the field of human vision, a human need at least two or three seconds to perceive a complex image [16]. Thus we used in this step the temporal redundancy which aims at filtering definitively the obtained regions.

In our work, we consider a temporal window of size N frames (N=50 ~ 2 seconds in video PAL format) in order to determine the persistence of the regions resulting from the candidates regions localization process, we apply the same technique to the frame reference (frame which just precedes that where the text appears for the first time) and each frames within this window with a shift of K frames. Each region having the same size, the same density and the same position throughout the temporal window will increment the duration of persistence $T_p$ by K. Thus, the regions having a value of $T_p$ equal to the size of the window N are kept whereas the others are ignored. Figure 5 illustrate such process.

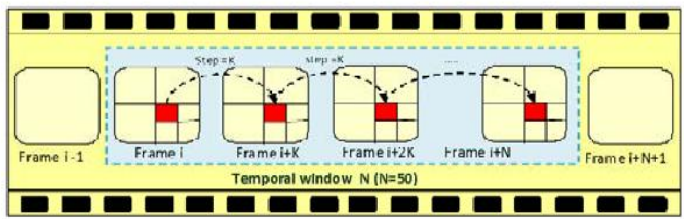

Figure 5. Temporal redundancy filtering process with a window of size N=50 and K=2

## III. EXPERIMENTS

Our experiments are conducted on sixteen video sequences captured at 25 frames per second with a 352x288 frame size and using a Pinnacle PCTV-sat card. These sequences (690.000 frames) are portions of news broadcasts, sports program and fiction movies. Within the test database text appears in a variety of colors, sizes, fonts and background complexity.

All static regions distinguishable as text by humans were included in the experimentation process. Closely spaced words lying along the same alignment were considered to belong to the same text instance. Test data contains a total of 882 temporally unique artificial text instances.

Our evaluation is based on the number of text regions, including region recall region precision and region false alarm.

$$Recall = \frac{Number\ of\ correted\ detected\ text\ regions}{Number\ of\ detected\ text\ regions} \quad (4)$$

$$Precision = \frac{Number\ of\ correted\ detected\ text\ regions}{Number\ of\ all\ ground\_truth\ text\ regions} \quad (5)$$

$$False\ alarm = \frac{Number\ of\ wrongly\ detected\ text\ regions}{Number\ of\ detected\ text\ regions} \quad (6)$$

Recall rate evaluates how many percents of the detected videotext regions are correct. Precision rate evaluates how many percents of all ground-truth video text regions are correctly detected. False alarm rate evaluates how many percents of the detected videotext regions are wrong as defined in (4), (5) and (6).

The overall results of our method were 89,01 % Recall, 88,05 % Precision and 11,95% as false alarm. These results show that our technique is efficient, capable to locate text regions with different character sizes and styles, even in case of texts occurring within complex image background and that lay on the boundary of the video frame. Besides, our algorithm has also the high speed advantage with 2 frame pair per second in the frame size of 352 x 288 which suitable for videotext detection. Furthermore, we find the detected text regions are more precise, where text strokes occupy more area. But we have some cases of fails mainly when a short text contains strokes belonging to terminal blocks having different parent blocks.

## IV. CONCLUSION

In this paper, we proposed an approach for video text detection, localization and extraction. This approach operates in two manly steps. The first is the candidate text regions the second is the filtering process. Within the first step we identify the starting frames of our algorithm by detecting the appearance of potential text objects. To localize these potential text regions, we use a split and merge processes applied on each binaries edge frames pair within video sequences. The resulting extracted regions are then filtered in order to get the effective one. Moreover, many applications can be derived from this automatic text locating technique. For instance, the automatic video summarization and the automatic generation of video content table enabling further video browsing and retrieval. In this context we developed a new prototype called "AViTExt: Automatic Video Text Extraction". This prototype can be used to browse video sequences by text images regions extracted automatically according to our proposed approach.

Although the effectiveness of our technique for detection of still text within video sequences, which give the most pertinent information about the content, we focuses our future work, firstly, on the determination of splitting and merging criteria which should be adaptive in step with the complexity of the background of the video frames and secondly on the integration of the video grammar as an interesting way to improve the performance of our approach and to guide the identification and the recognition of text regions. Consequently, our prototype will enable a full text search within a video database.